\documentclass[preprint,double-spaced,showpacs,amsmath,amssymb]{revtex4}
\usepackage{graphicx}
\usepackage{dcolumn}
\usepackage{bm}
\newcommand{\beq}{\begin{equation}}
\newcommand{\eeq}{\end{equation}}
\newcommand{\bea}{\begin{eqnarray}}
\newcommand{\eea}{\end{eqnarray}}
\newcommand{\eps}{\varepsilon}

\begin{document}

\author{S. Kamerdzhiev}
\affiliation{Institute of Physics and Power Engineering, 249033 Obninsk, Russia,}
\author{E.E. Saperstein}
\affiliation{Kurchatov Institute, 123182 Moscow.}

\title{Interaction of the single-particle and collective degrees of freedom
       in non-magic nuclei: the role of phonon tadpole terms}


\begin{abstract}
A method of a consistent consideration of the phonon contributions
to mass and gap operators in non-magic nuclei is developed in the
so-called $g^{2}$ approximation, where $g$ is the low-lying phonon
creation amplitude. It includes simultaneous accounting for both
the usual non-local terms and the phonon tadpole ones. The
relations which allow the tadpoles to be calculated without any
new parameters are derived. As an application of the results, the
role of the phonon tadpoles in the single-particle strength
distribution and in the single-particle energies and gap values
has been considered. Relation to the problem of the surface nature
of pairing is discussed.
\end{abstract}
\maketitle

\section{introduction}
In the last two decades, the progress was achieved in the many-body
nuclear theory \cite{AB,Schuck} in going beyond the standard RPA
or the Theory of  Finite Fermi Systems (TFFS) by means of
accounting for coupling of the single-particle degrees of freedom
with the low-lying  collective excitations (``phonons''). For
magic nuclei Refs.~\cite{KhS,bbb1983,ETFFS2004,litvaring} could be
cited, and  for nuclei with pairing these are
Refs.~\cite{baranco1999,avekaevlet1999,milan2,avekaev1999,tselyaev2007}.
For the first studies in this field the use of the
phenomenological Saxon-Woods mean field is typical. In this case,
a double set of the phenomenological parameters is necessary, the
first one for the effective force and the second one for the mean
field. Note that in Ref.~\cite{KhS} the main terms of the mean
field potential well, i.e. the ones without the phonon
corrections, were found in a self-consistent way, within the
self-consistent version of the TFFS. However, the phonon
characteristics were calculated with the use of the Saxon-Woods
potential, thereby the double set of the parameters was used, too.

The characteristic feature of  the  modern developments in this
field is the complete refusal from the phenomenological  mean
field when both the field and the phonon characteristics are
calculated self-consistently, with the use of only one set of
parameters for the  forces to calculate the mean field.
 The HFB calculations with the
Skyrme forces \cite{sarchi,ave2007} should be cited here and the
ones, for magic nuclei only, within the relativistic mean field
theory \cite{litvaring,litvaringvret2007}. Article
\cite{litvaring} is an expressive example. Here, in the
framework of the old dynamic scheme of \cite{ringwerner1973}, the
Saxon-Woods  mean field was changed by the relativistic mean
field.  In both studies the authors solved the Dyson equation for
the mass operator in the so-called g$^{2}$ approximation, g being
the  phonon-particle coupling  amplitude. Properties of the odd
nuclei nearby $^{208}$Pb were considered, including
the single-particle strength distribution and  characteristics of the
 single-particle levels.

A consistent generalization of
 the g$^{2}$ approximation to the nuclei with pairing has been
performed in \cite{avekaev1999}.
 There Eliashberg's approach \cite{eliashberg1961},
which was developed originally for superconductivity in the solid
state physics, has been used and the realistic calculations have
been performed for $^{121}$Sn and $^{123}$Sn nuclei.
 In Refs.~\cite{avekaev1999,avekaevlet1999} the problem of evaluating the
 phonon contribution to the single-particle
energies and gap values was consistently formulated and the
equations obtained were solved.

The problem of the phonon contribution to the nuclear gap value
is, probably, one of the most interesting at present because it is
in a close relation to the problem of nature of the nuclear
superfluidity. Note that it is also of great interest for the
astrophysics, see, e.g., Refs.~\cite{milan1,Baldo05}. As to atomic
nuclei themselves, the old problem exists  whether the nuclear
superfluidity has a volume or surface nature. In the latter case,
a question remains in what extent it appears due to a special form
of the initial pairing force or to an additional
contribution induced by exchange with the collective low-energy
surface phonons. The calculations in \cite{avekaevlet1999} have shown
that we deal with an intermediate case: the phonon contribution to
the gap value $\Delta$ for the tin region is about 30\% of the
experimental gap value, the rest of about 70\% being due to the
``main'' pairing force.

Simultaneously and independently, the problem of nuclear pairing
was attacked by the Milan group. They combined an {\it ab initio}
calculation of the gap starting with the Argonne v$_{14}$
NN-force, Refs.~\cite{milan0,milan2}, with evaluating the phonon
contribution to $\Delta$, Refs.~\cite{baranco1999,milan2}. Their
conclusion  was that about 50\% of the gap value for the tin region
appears due to the initial NN-force and the other 50\% due to
the phonon contribution. Note that the alternative {\it ab initio}
calculation of the gap was carried out recently in
Refs.~\cite{Pankratov1} (Paris NN-force) and \cite{Pankratov2}
(Argonne v$_{18}$ NN-force) on the basis of the method developed in
\cite{zverevsap2004}. Omitting the  discussion of the two methods of
 solving the {\it ab initio} gap equation, note only that a close
agreement with the experimental value of $\Delta$ was obtained in
Refs.~\cite{Pankratov1}, \cite{Pankratov2}, leaving a room
of not more than 20\%for the
phonon corrections. It agrees qualitatively
with the results  \cite{avekaev1999,avekaevlet1999}, but
 significantly contradicts those of
Refs.~\cite{baranco1999,milan2}. It is worth mentioning that in
the cited papers by the Milan group the phonon contribution to the
gap of  Ca and Ti isotopes reaches almost 100\%. Thus, a
conflicting situation  remains in the problem under discussion and
new investigations are required to clear up the situation.

However, in all of  the above-mentioned papers, and in many other
studies dealing with evaluation of the low-lying phonon
contribution to nuclear characteristics, a term was lost, which is
of the  same $g^{2}$ order as the usual non-local pole  term of
the mass operator taken into account. We mean the so-called
tadpole term. It was evaluated firstly in the pioneering paper by
V.A. Khodel \cite{khod1} and named as the local term. Now we
name it, in accordance with the particle physics terminology, as
the tadpole diagram. This approach is closely related to the
interpretation of the low-lying surface phonons as members of the
Goldstone branch which appear  due to spontaneous breaking
 of the translation invariance in nuclei. The ghost $1^-$-state with
the frequency $\omega_1=0$ is the head of this branch, being the
exact solution of the self-consistency equation of the TFFS.
Therefore, for the simplest version of the TFFS effective force,
the corresponding eigenfunction is $g_1=\frac{dU}{dr}$, where
$U(r)$ is the mean field potential. The TFFS equations for the
natural parity excitations will yield small excitation energies
$\omega_L$ and eigenfunctions $g_L$ close to $g_1$, provided the
self-consistency relation is fulfilled. The main idea of
Ref.~\cite{khod1} was to develop such a scheme of evaluating the
surface phonon corrections to nuclear characteristics that they
should vanish for the ghost phonon.
 As it turned out, that  is impossible without taking into
account the tadpole term.

 For magic nuclei the tadpole term
was considered in detail  in Ref.~\cite{KhS}, see also references
therein. As it turned out, the tadpole contribution  to the
single-particle energies, splitting of the particle-vibration
multiplets and other properties of magic nuclei and their odd
neighbors are, as a rule, important and are often  of the opposite
sign as compared with the the usual non-local terms. Development
of an analogous approach for non-magic nuclei is a problem of
great interest.

The present paper is the  first step in this direction. In Section
II we derive a closed set of equations for the tadpole operators
in nuclei with pairing, which doesn't contain any new parameters
in addition to those used for calculating the mean-field mass
operators. In Section III we consider possible approximations to
make this set more handy. Closed and transparent relations for the
particle-hole (ph) and particle-particle (pp) tadpole terms are
obtained. As an application, modifications of the results of
Refs.~\cite{avekaev1999,avekaevlet1999} for the single-particle
strength distribution and single-particle energies and gap values
of superfluid nuclei due to inclusion of the tadpoles are carried
out. The relevance of the tadpole terms to the problem of the
nature of nuclear pairing is discussed.

\section{Phonon corrections to the mass and gap operators}

In this paper, we deal with the nuclei with weak phonon-particle
coupling when the phonon admixture to the one-particle degrees of
freedom could be accounted for within a perturbation theory
scheme. More specifically, the quantity
\begin{equation}
 \alpha = \frac {\bar g_L^2} {(2j+1)\omega_L^2} \label{alp}
 \end{equation}
plays the role  of the perturbation parameter, and it should be
small. Here $\bar g_L$ is the average value of the matrix element
of the $L$-phonon creation amplitude at the Fermi surface,
$\omega_L$ is its excitation energy, and $j$ is
 a typical value
of the single-particle angular momentum ($j \propto A^{1/3}$). In
other words, the coupling strength should be not too high and the
excitation energy not too low. We call this as the $g^{2}$
approximation for mass operators.

Such a situation takes place in magic and semi-magic nuclei. There
 is no pairing at all in the first case and partially, in
the magic subsystem, in the second one. For the magic nuclei, the
problem under consideration was consistently solved in \cite{KhS},
see also references therein. Our aim is to develop a similar
approach for semi-magic nuclei, with  pairing in
the non-magic subsystem taken into account.

In the general case of nuclei with pairing  it is necessary
 to consider  four one-particle generalized Green functions, i.e.
 two Green functions,  $G$ and $G^{h}$,
 and two Gor'kov functions $F^{(1)}$ and $F^{(2)}$. In addition
to two   normal  mass operators, $\Sigma$ and $\Sigma^{h}$,
their two  anomalous counterparts appear. In the textbook
\cite{Landau} they are denoted as $\Sigma_{02}$ and $\Sigma_{20}$,
in Refs.~\cite{avekaevlet1999,avekaev1999} as $\Sigma^{(1)}$ and
$\Sigma^{(2)}$. Here we use the notation  \cite{AB}, where they
are denoted as $\Delta^{(1)}$ and $\Delta^{(2)}$. Therefore we
will often  name them as the gap operators. Sometimes we use the
term ''mass operators`` for all the four quantities under discussion.

 The main part of the mass operators is determined by the mean field
 contribution. In the $g^{2}$ approximation, for the single-particle
 energy ${\eps}_{\lambda}$ and the average gap value ${\Delta}_{\lambda}$ we have:
 \beq
{\eps}_{\lambda} = \Sigma_{\lambda \lambda} =
{\eps}^{(0)}_{\lambda} + \delta ^{(2)} \Sigma _{\lambda
\lambda}(\eps_{\lambda}),
 \hspace{1cm}
 \Delta_{\lambda} =  \Delta_{\lambda \lambda} = \Delta_{\lambda \lambda}^{(0)}
  +  \delta ^{(2)}\Delta_{\lambda \lambda}(\eps_{\lambda}) \label{phon_cor},
  \eeq
with the obvious notation. In general, the main, mean field, parts
of $\Sigma$ and $\Delta$, in the approach discussed, are supposed
to be calculated within a self-consistent method, say, the HFB one
or the self-consistent TFFS.  However, in practice the
phenomenological Woods-Saxon potential and pairing force are often
used. It should be emphasized  that in the latter case it is
necessary to use the so-called ``refinement'' procedure
\cite{avekaev1999,ETFFS2004} in order  to avoid a double counting
of phonon contribution into the terms $ {\eps}^{(0)}_{\lambda}$
 and the gap $ {\Delta}^{(0)}_{\lambda}$ and to extract the ``refined'' values
from the phenomenological ones. In principle, analogous
precautions should be made in the case of the self-consistent
calculation with the use of phenomenological forces, too.
Indeed, the force parameters should be chosen in such a way that
the total expressions (\ref{phon_cor}) (not the ''zero`` ones) would
reproduce the experimental values. Note that this very idea was
utilized in \cite{KhS}.

  With the use of  the $g^{2}$ approximation one can write down
the phonon corrections, e.g., to the  operators $\Sigma$ and
$\Delta^{(1)}$, see Fig.1 and Fig.2, as follows:

\begin{figure}[]
{\includegraphics [height=20mm]{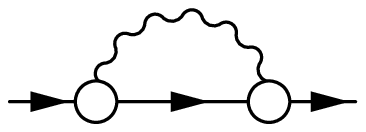}} \hspace{0.5cm} +
\hspace{0.5cm} {\includegraphics [height=20mm]{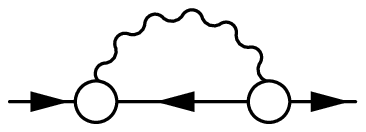}}
\hspace{0.5cm} + \hspace{0.5cm} {\includegraphics
[height=20mm]{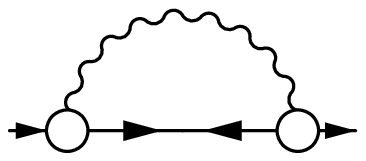}} \hspace{0.5cm} + \hspace{0.5cm}
{\includegraphics [height=20mm]{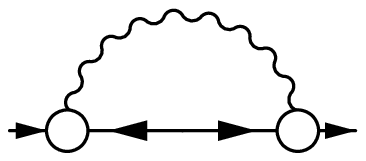}} \hspace{0.5cm} +
\hspace{0.5cm} {\includegraphics [height=20mm]{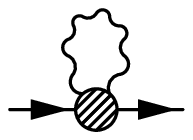}}
\caption{Phonon $g^{2}$ corrections to the mass operator
$\Sigma(\eps)$ in non-magic nuclei (general case).}
\end{figure}

  \begin{figure}[]
{\includegraphics [height=20mm]{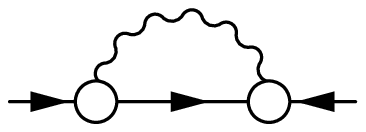}} \hspace{0.5cm} +
\hspace{0.5cm} {\includegraphics [height=20mm]{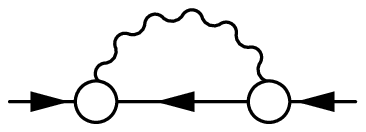}}
\hspace{0.5cm} + \hspace{0.5cm} {\includegraphics
[height=20mm]{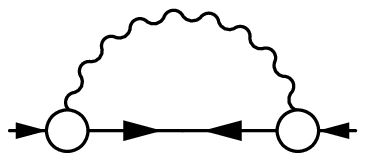}} \hspace{0.5cm} + \hspace{0.5cm}
{\includegraphics [height=20mm]{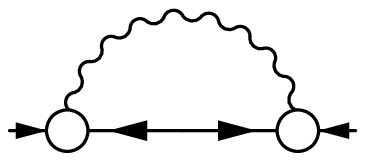}} \hspace{0.5cm} +
\hspace{0.5cm} {\includegraphics [height=20mm]{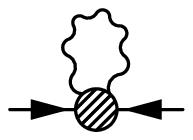}}
\caption{Phonon $g^{2}$ corrections to the gap operator
$\Delta^{(1)}(\eps)$ in non-magic nuclei (general case).}
\end{figure}

 \beq
 \delta ^{(2)}\Sigma (\eps) =
  {M}(\eps)  + K^{ph}\,,
\label{mig}
 \eeq
 \beq
 \delta ^{(2)}\Delta^{(1)} (\eps) =
{M}^{(1)}(\eps) + K^{pp}\,.
 \label{mig1}
 \eeq
The set of diagrams and the relation for $\delta
^{(2)}\Delta^{(2)}$ are  absolutely similar to Fig.2 and
Eq.~(\ref{mig1}). In Fig. 1 and Fig. 2 the empty circles denote
the phonon creation amplitudes (vertexes). To distinguish between
the $g$- and $d$-vertexes it is necessary just to look  at the
direction of ingoing and outgoing arrows. In the case of one
ingoing and one outgoing arrow we deal with the $g$- (or $g^h$-)
vertex. If there are two ingoing arrows, we deal with the
$d^{(1)}$-vertex, two outgoing arrows mean the $d^{(2)}$-vertex. The
terms with two phonon creation amplitudes are the energy dependent
 non-local operators, the sums of them being denoted as $M$
and $ M^{(1)}$. The last terms are the corresponding ph- and
pp-phonon tadpoles. The essential property of these tadpoles is
that they do not depend on the energy $\eps$.

 The explicit expressions for the phonon
tadpoles K$^{ph}$ and K $^{pp}$ are given by \beq K^{ph} = \int
\frac{d\omega}{2\pi i} D(\omega)\delta g (\omega)\,, \label{Kph}
\eeq \beq K^{pp} = \int \frac{d\omega}{2\pi i} D(\omega)\delta d
^{(1)}(\omega) \,,\label{Kpp} \eeq where D is the phonon Green
function  and $\delta g$ and $\delta d^{(1)}$ are the changes of
the ph- and pp-phonon creation amplitudes in the external field of
 another phonon with the same quantum numbers as the one  whose
contribution we analyze.

\subsection{Magic nuclei}

To begin with, let us first outline briefly the method for magic
nuclei, following to \cite{KhS}. In this case there  is no pairing
 and only the first and the last terms in Fig. 1 for the corrections
to the mass operator $\Sigma$ survive. The first term is the usual
pole diagram, where the Green function G, of course, does not
contain pairing effects. The last term means the sum of all the
irreducible diagrams.  In the problem under consideration, as it
was mentioned in the Introduction, this sum was evaluated firstly
in Ref.~\cite{khod1}. We name it, in accordance with the particle
physics terminology, as the tadpole diagram.

The second order in $g_L$ correction to the mass operator $\Sigma$
reads \beq \delta_{LL}^{(2)}\Sigma(\eps) = \int \frac
{d\omega}{2\pi i} S_{LL}(\eps,\omega) D_L(\omega), \label{d2Sig}
\eeq where $D_L$ stands for the $L$-phonon $D$-function and
$S_{LL}$ is the phonon-particle scattering amplitude. As usual,
the symbolic multiplication means the integration over
intermediate coordinates and summation over the spin variables. In
accordance with Fig.1, $S_{LL}$ is the sum \beq
 S_{LL} = g_L G g_L + \delta_L{g_L}\,,
 \label{M_LL}
 \eeq
where $\delta_L{g_L}$ is the tadpole term. Obviously, it is of the
second order correction to $\Sigma$, as far as the first order
correction to $\Sigma$ is the particle-phonon interaction
amplitude, $g_L=\delta_L \Sigma$. According to the recipe of
\cite{KhS}, it can be found by the direct variation of the
equation for the vertex $g_L$, \beq g_L = {\cal F} A g_L,
\label{eq_g} \eeq where ${\cal F}$ is the Landau--Migdal effective
NN-interaction amplitude \cite{AB} and $A=GG$ is the particle-hole
propagator. After the variation of Eq.(\ref{eq_g}) over the phonon
creation amplitude one obtains: \beq \delta_L g_L = ({\delta_L
\cal F}) A g_L + {\cal F} (\delta_L A) g_L + {\cal F} A \;\delta_L
g_L. \label{eq_dg} \eeq This is an integral equation for the
quantity $\delta_L g_L$ with the   inhomogeneous term \beq m_{LL}
= (\delta_L {\cal F}) A g_L + {\cal F} (\delta_L A) g_L.
\label{m_sm} \eeq

The procedure of finding the second term of the inhomogeneous term
is quite obvious. The direct variation of the particle-hole
propagator yields:
 \beq
 (\delta_L A) = 2 G(\delta_L G) = 2 G G g_L G. \label{dA}
 \eeq

 Up to now, we have used a symbolic notation. To obtain the explicit
relations, let us, for simplicity, suppose that the mass operator
$\Sigma$ is momentum independent. In this case, we have  for the
first order correction \beq \delta_{LM}^{(1)}\Sigma({\bf r }) =
g_{LM}({\bf r}) = g_L(r) Y_{LM}({\bf n}). \label{gL} \eeq Note
that within the Bohr--Mottelson (BM) liquid drop model one has
\beq g_L^{\rm BM}(r)=\alpha_L \frac {dU(r)}{dr}, \label{gBM} \eeq
where $U(r)$ stands for the nuclear mean field potential and
$\alpha_L$ is the constant which determines the amplitude of the
surface vibration. As it was demonstrated in \cite{KhS}, the
direct solution of the RPA-like equation (\ref{eq_g}) for a
low-lying excitation in an even-even nucleus is very close to the
BM model prescription (\ref{gBM}). Therefore,  for qualitative
estimations,  one can use this simplified form of the
particle-phonon vertex.

 To obtain the second term in (\ref{m_sm}) we could fold the quantity
(\ref{dA}) with $g_L$. Let us introduce the notation $T=\delta_L A
g_L$. In the explicit form, with the help of (\ref{dA}), we
obtain \bea T_{LM_1LM_2}({\bf r},\omega) & = &\int \frac
{d\eps}{2\pi i} d {\bf r}_1 d {\bf r}_2 G({\bf r},{\bf r}_1;\eps) g_{LM_1}({\bf r}_1)\notag \\
&\times& \left[G({\bf r}_1,{\bf r}_2;\eps - \omega) + G({\bf
r}_1,{\bf r}_2;\eps + \omega) \right]  g_{LM_2}({\bf r}_2) G({\bf
r}_2,{\bf r};\eps). \label{TLL}
 \eea

The problem of finding the first term of (\ref{m_sm}) looks less
obvious. To find it,  an ansatz was used in \cite{KhS} based on
the density dependence of the Landau--Migdal amplitude $\cal{F}$,
\beq \delta_L {\cal F} = \frac { \delta {\cal F}} {\delta \rho }
\delta_L \rho,  \label{dF} \eeq where $\delta_L \rho$ is the
transition  density associated with the $L$-phonon excitation. It
obeys the relation \beq {\delta_L \rho } = A g_L. \label{drho}
\eeq In the approximation similar to (\ref{gBM}),
 we have
\beq ({\delta_L \rho })^{\rm BM}(r)=\alpha_L \frac {d\rho(r)}{dr}.
\label{rhoBM} \eeq

 Thus, the equation for the ph-tadpole in magic nuclei reads
  \beq K^{ph} = \delta_{L} {\cal F} A g D +  {\cal F}
(\delta_L A) g_L D + {\cal F} A K^{ph}. \label{eq_K_simb} \eeq In
the graphic form this equation is shown in Fig. 3.

\begin{figure}[]
  {\includegraphics [height=20mm]{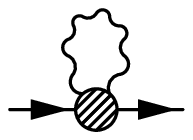}} =
  {\includegraphics [height=20mm]{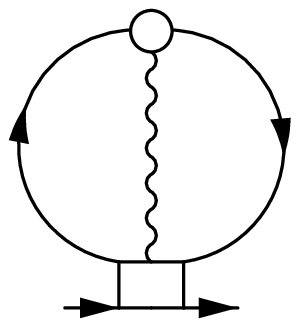}}
 + {\includegraphics [height=20mm]{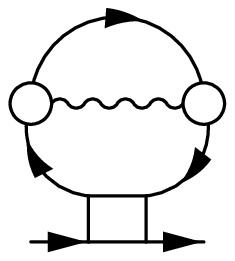}}
+ {\includegraphics [height=20mm]{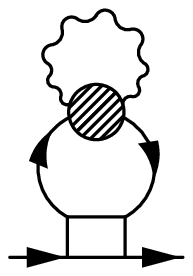}} \caption{Equation
for the tadpole in magic nuclei}
\end{figure}

\subsection{Non-magic nuclei}
As it was discussed above, in the general case of nuclei with
pairing, in addition to the one-particle Green function $G$, two
Gor'kov functions $F^{(1,2)}$
 enter the TFFS  relations, and two gap functions
$\Delta^{(1,2)}$ appear in addition to the mass operator $\Sigma$.
They are related to each others, \beq \Delta^{(1,2)}={\cal
F}^{\xi} F^{(1,2)}, \label{Del12} \eeq in terms of the
 interaction amplitude ${\cal F}^{\xi}$ irreducible
  in the particle-particle channel.
 It should be noted that hereafter we mean that  the
Green function $G$  takes the superfluidity effects
into account.

In the systems with pairing, Eq.(\ref{eq_g}) for the phonon-particle
vertex is generalized and two new  amplitudes appear \cite{AB}:
\beq d^{(1,2)}_{LM}({\bf r}) =\delta_{LM}\Delta^{(1,2)}({\bf r })
= d^{(1,2)} _L(r) Y_{LM}({\bf n}). \label{d12} \eeq Let us omit
for a while the subscripts and upper indices. As far as the
amplitude ${\cal F}^{\xi}$ in the TFFS  is considered to be
density dependent \cite{SapTr,ZverSap,Fay}, two terms in
(\ref{d12}) appear: \beq d ={\cal F}^{\xi} \delta F + \delta {\cal
F}^{\xi} F. \label{two_d}\eeq
It is worth pointing out  that usually
 the first term in (\ref{two_d}) is only taken  into account. V.A.
Khodel \cite{khod2} was, evidently, the first who turned  attention to the second
one. It is natural to use for it the ansatz analogous
to (\ref{dF}): \beq \delta_L {\cal F}^{\xi} = \frac { \delta {\cal
F}^{\xi}} {\delta \rho } \delta_L \rho,  \label{dFxi} \eeq but
now, due to pairing effects, the transition density obeys the
equation which is more complicated than Eq.~(\ref{drho}). It will
be written  down below.

To introduce the standard TFFS  notation, let us omit for a while
the second term in Eq.~(\ref{two_d}). As far as we deal with the
low-lying excitations of natural parity, contributions of
spin-dependent forces could be neglected in the equations for g
and g$^h$ \cite{AB}. As the result, the relation $g^h=g$ is valid.
In this case, the quantities $g_L$ and $d^{(1,2)}$ obey the set of
equations \cite{AB}, which could be written in the form similar to
(\ref{eq_g}),
 \beq
 {\hat
g}={\hat {\cal F}}  {\hat A} {\hat g}, \label{eq_gex} \eeq but now
all the ingredients of (\ref{eq_gex}) are matrices: \beq {\hat
g}=\left(\begin{array}{c}g
\\d^{(1)}\\d^{(2)}\end{array}\right)\,,
\label{ds} \eeq

\beq {\hat {\cal F}}=\left(\begin{array}{ccc} {\cal F} &0
&0\\
0&{\cal F}^\xi  &0\\
0&0& {\cal F}^\xi \end{array}\right), \label{Fs} \eeq

\beq {\hat A}=\left(\begin{array}{ccc} {\cal L} &{\cal M}^{(1)}
&{\cal M}^{(2)}\\
 {\cal O}&{\cal N}^{(1)} &{\cal N}^{(2)}\\\tilde{\cal O}&\tilde{\cal N}^{(1)} &
 \tilde{\cal N}^{(2)}
\end{array}\right)\,.
\label{As} \eeq Here  ${\cal L},\; {\cal M}^{(1)}$, and so on,
denote  integrals over $\eps$ of different double products of the
Green function $G(\eps)$ and Gor'kov functions, $F^{(1)}(\eps)$
and $F^{(2)}(\eps)$. They could be found in \cite{AB} and we write
down now explicitly only two of them, \beq {\cal
L}(\omega)=\int\frac{d\eps}{2\pi
i}\left[G(\eps)G(\eps+\omega)-F^{(1)} (\eps)F^{(2)}
(\eps+\omega)\right], \label{Ls} \eeq and \beq {\cal O}(\omega)=
-\int \frac{d\eps}{2\pi i} \left[G(\eps) F(\eps+\omega) + F(\eps)
G(-\eps-\omega)\right]. \label{Os} \eeq

Let us come back to the second term of (\ref{dFxi}). With the help of
the above short notation the transition density could be written
in a compact form similar to (\ref{drho}): \beq {\delta \rho } =
\sum_i A_{1i} g_i. \label{drhos} \eeq Using this relation, we may
express the term under consideration in terms of the ``generalized
vertex function'' $\hat g$: \beq \delta {\cal F}^{\xi} = \frac {
\delta {\cal F}^{\xi}} {\delta \rho } \sum_i A_{1i} g_i.
\label{dFxi1} \eeq

By substituting this relation to (\ref{dFxi}) we find that the
general structure of Eq.~(\ref{eq_gex}) remains valid, but now the
``interaction matrix'' $\hat {\cal F}$ becomes more complicated,
in particular, non-diagonal. To be more exact, the first line of
(\ref{Fs}) remains unchanged, but new non-diagonal terms appear in
two other lines. To simplify their explicit form, let us use the
approximation for the effective pairing interaction amplitude
${\cal F}^{\xi}$ which is usually utilized in the TFFS (e.g., see
\cite{Fay}). Namely, it is considered as an energy independent
delta-force with a density dependent strength ${\cal
F}^{\xi}(\rho({\bf r}))$. In this case, the second term of
(\ref{two_d}) is reduced to \beq \delta {\cal F}^{\xi} F = \frac {
d{\cal F}^{\xi}} {d \rho } \; {\delta \rho } ({\bf r}) \;\chi({\bf
r}),\label{dFxichi} \eeq where \beq \chi({\bf r}) = \int \frac
{d\eps}{2\pi i}
 F(\eps,{\bf r},{\bf r})  \label{chi} \eeq is the anomalous
 density. Combining the above relations, one can readily find two
 new non-diagonal terms of the matrix $\hat {\cal F}$,
\beq {\cal F}_{21}={\cal F}_{31}= \frac { d {\cal F}^{\xi}} {d
\rho }\;\chi({\bf r}) . \label{F_21} \eeq

Thus, we obtain the matrix equation (\ref{eq_gex}) for  $\hat g$
in the general case where both terms of Eq.~(\ref{two_d}) are
taken into account. After variation of this equation over the
field of the surface $L$-phonon under consideration we find the
matrix equation for the tadpole term in a superfluid nucleus: \beq
\delta_L \hat g_L = (\delta_L \hat {\cal F}) \hat A \hat g_L +
\hat {\cal F} (\delta_L \hat A)\hat g_L +\hat {\cal F}\hat A
\;\delta_L\hat g_L. \label{eq_dg_ex} \eeq

In principle, this set of equations solves the problem of finding
the tadpole terms under discussion.
 In accordance with Eqs.~(\ref{mig}),(\ref{mig1}), they should
be obtained by folding the solutions of Eq.~(\ref{eq_dg_ex}) with the
phonon $D$-function. However, the explicit form of this equation
is quite cumbersome. The second term on the r.h.s. of
Eq.~(\ref{eq_dg_ex}) is the most complicated. To find $\delta
{\cal L}$ and variations of other components of the $\hat A$
matrix, one can use the well known expressions for $\delta G$,
$\delta G^h$, $\delta F^{(1,2)}$ \cite{AB}. In the result, one
obtains a lot of integrals of triple combinations of the Green and
Gor'kov functions of the type of Eq.~(\ref{TLL}). To obtain  more
handy relations, some approximations should be made.

\section{ Small $d$ approximation}

As it was discussed in the Introduction, the collectivity of the
low-lying ph-phonons, i.e. the surface vibrations, exceeds
significantly that of the pp-phonons, i.e. the  pairing
vibrations. Therefore, we concentrate here on the contributions to
the mass and gap operators of the phonons of the first type. In
this case, the $g_L$ component of the generalized vertex $\hat
g_L$ dominates in  Eqs.~(\ref{eq_gex}) and (\ref{eq_dg_ex}) and
the inequality  $g \gg d^{(1,2)}$ is valid. For this reason, we
can omit the terms with the pairing phonon creation
 amplitudes $d^{(1,2)}$ in these equations  and, correspondingly, in Figs. 1,2.
  In this approximation, the diagrams
 depicted in Figs. 4,5 should only be taken into account . In
 addition, we assume that we deal with the ``developed pairing''
 case when pairing properties of neighboring even-even nuclei
 should  be considered identical. In this case, we have
 $\Delta^{(1)}=\Delta^{(2)}=\Delta$ and $d^{(1)}(\omega)=\pm
 d^{(2)}(-\omega)=d(\omega)$\cite{AB}. The sign ``+'' takes place
 for the states of natural parity we consider.
 The appropriate explicit expressions for $M(\eps)$=$M^{h}(-\eps)$
 and $M^{(1)}(\eps)$=$M^{(2)}(\eps)$
 are given in \cite{avekaev1999}.

  \begin{figure}[]
{\includegraphics [height=20mm]{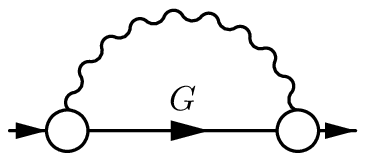}} \hspace{0.5cm} +
\hspace{0.5cm} {\includegraphics [height=20mm]{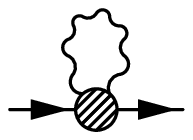}}
\caption{Phonon $g^{2}$corrections to the mass operator
$\Sigma(\eps)$ in non-magic nuclei in the small $d$  approximation
 (the Green function G contains pairing effects).}
\end{figure}

    \begin{figure}[]
{\includegraphics [height=20mm]{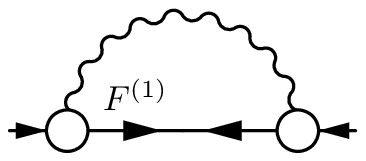}}  \hspace{0.5cm} +
\hspace{0.5cm} {\includegraphics [height=20mm]{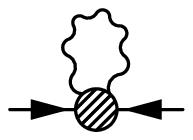}}
\caption{Phonon $g^{2}$corrections to the gap operator
$\Delta^{(1)}(\eps)$ in non-magic nuclei in the small $d$
approximation.}
\end{figure}

 In the approximation under consideration, instead of the set (\ref{eq_gex}),
we obtain the closed equation for
 $g$,
 \beq
 g = {\cal F }{\cal L} g\,,
 \label{g_pair}
 \eeq
 and  the closed expression for the $d$-vertex in terms of $g$:
  \beq d(\omega) = \left( {\cal F}_{21}
  {\cal L}(\omega) +   {\cal F}^{\xi}{\cal O}(\omega)\right) g.  \label{d_pair}
 \eeq
Sometimes the initial equation (\ref{two_d}) for the $d$-vertex is
more convenient. In the small $d$ approximation, it reads:
  \beq d(\omega) ={\cal F}^{\xi}{\cal O}(\omega) g + \delta{\cal F}^{\xi} F\,.  \label{d0_pair}
 \eeq

The set of Eqs.~(\ref{eq_dg_ex}) for the tadpole terms is also
simplified. It could be obtained either  by omitting terms containing
$d^{(1)},d^{(2)}$ in (\ref{eq_dg_ex}) or by the direct variation
of Eqs. (\ref{g_pair}) and (\ref{d_pair}). In the result, one
receives: \beq \delta g =
 \delta {\cal F} {\cal L} g + {\cal F} \delta {\cal L} g +
{\cal F} {\cal L} \delta g, \label{delta_g} \eeq \beq \delta d =
\delta {\cal F}_{21} {\cal L} g + {\cal F}_{21} \delta {\cal L} g
+ \delta {\cal F}^{\xi} {\cal O} g + {\cal F}^{\xi} \delta {\cal
O} g
 + \left( {\cal F}_{21} {\cal L}+{\cal F}^{\xi} {\cal O}\right) \delta
 g\,.
 \label{delta_d}
 \eeq

 Thus, we obtained the integral equation (\ref{delta_g})
 for $\delta g$ with the inhomogeneous term, which is similar to the $m_{LL}$
 term  (\ref{m_sm}) for magic nuclei,
 and expression (\ref{delta_d}) for $\delta d$. The latter
 contains $\delta g$ and four terms,  which are analogous to the
 inhomogeneous terms of the equation for the $\delta g$. Let us
 consider them in  detail.

An alternative relation for $\delta d$ could be found by variation
of Eq.~(\ref{d0_pair}). It is as follows: \beq \delta d = \delta
{\cal F}^{\xi} {\cal O} g + {\cal F}^{\xi} \delta {\cal O} g +
\delta{\cal F}^{\xi} \delta F + (\delta^2 {\cal F}^{\xi}) F
 + {\cal F}^{\xi} {\cal O} \delta
 g\,.
 \label{delta_d0}
 \eeq

To obtain the quantities $\delta {\cal L}$ and $\delta {\cal O}$
in the equations discussed above in the explicit form, one should
variate the propagators ${\cal L}$ and ${\cal O}$: \beq \delta
{\cal L} = \delta (G G - F^{(1)}F^{(2)})\,, \label{delta_L} \eeq
\beq
 \delta {\cal O} =  \delta (G F^{(1)} + F^{(1)}G^{h}),\label{delta_O}
\eeq and use the small $d$ approximation, omitting the terms with
$d^{(1)},d^{(2)}$ in the general expressions for the variation of
the Green functions \cite{AB}. One obtains: \beq
 \delta G = G g G -F^{(1)}g F^{(2)} ,\hspace{1cm}
 \delta G^{h} = G^{h} g G^{h} - F^{(2)} g F^{(1)}\,,
 \label{d_G_G(h)pair}
 \eeq
 \beq
 \delta F^{(1)} = G g F^{(1)} + F^{(1)} g^{h} G^{h} , \hspace{1cm}
 \delta F^{(2)} = F^{(2)} g G + G^{h} g^{h} F^{(2)}\,.
 \label{d_F1_F2pair}
 \eeq
For a time, we come back to the notation $F^{(1)}, F^{(2)} , G^{h}$
to avoid an explicit specification of the energy variables in the
integrals similar to that in Eq.~(\ref{TLL}), which appear after
folding expressions (\ref{delta_L}) and (\ref{delta_O}) with $g$.
They could be obtained by combining
Eqs.(\ref{delta_L})-(\ref{d_F1_F2pair}) and, in the symbolic form,
are as follows:
 \bea
 \delta {\cal L} g =& g(G G G - F^{(1)} F^{(2)} G + G G G - G F^{(1)} F^{(2)}
 -\notag \\
 &G F^{(1)} F^{(2)} -  F^{(1)} G F^{(2)} - F^{(1)} F^{(2)} G - F^{(1)} G^{h} F^{(2)}) g\,, \label{d_Lg}
 \eea
and
 \bea
 -\delta {\cal O} g = g(GGF^{(1)} - F^{(1)}F^{(2)}F^{(1)} + GGF^{(1)} + GF^{(1)}G^{h} +\notag \\
                        GF^{(1)}G^{h} + F^{(1)}G^{h}G^{h} + F^{(1)}G^{h}G^{h} -
                        F^{(1)}F^{(2)}F^{(1)})g\,.
 \label{d_Og}
 \eea

 Thus, even for the simplified case under consideration, we obtained eight
terms for $\delta {\cal L} g$ instead of the one in
Eq.~(\ref{eq_dg}). In addition to them,  eight new terms for
$\delta {\cal O} g$ appear in the expression for $\delta d$. The
explicit form of each integral entering
Eqs.~(\ref{d_Lg}),(\ref{d_Og}) is  similar to that of
Eq.~(\ref{TLL}).

 \subsection{Final relationships for the tadpoles}
For magic nuclei, equation (\ref{eq_K_simb}) for the tadpole
term was solved in the coordinate representation \cite{KhS}. Even
in this case the procedure turned out to be quite cumbersome. In
principle, this method could be generalized to the systems with
pairing, using the coordinate representation for the Green and
Gor'kov functions \cite{fayansbelyaev}.  However, as it is clear
from the above formulas, in this case it will be much more
complicated. For this reason, we  prefer to use  the
representation of the single-particle wave functions, the
so-called $\lambda$-representation. To make the final equations
for the tadpoles more transparent, we also use the diagonal in
$\lambda$ approximation. The matter is that the set $\{\lambda\}$
is chosen in such a way that the mean field mass operator
$\Sigma^{(0)}$ and the corresponding Green function $G^{(0)}$ are
diagonal in $\lambda$. We use the  approximation supposing that the mean
field gap function $\Delta^{(0)}$, the Green function $G$ with
pairing and Gor'kov functions $F^{(1,2)}$ are  also  diagonal in
$\lambda$ \cite{AB}. In this approximation, ${\cal L},{\cal O}$
and other two-particle propagators contain two
$\lambda$-subscripts, ${\cal L}_{\lambda_1\lambda_2}$ and so on,
whereas in the general case we have ${\cal
L}_{\lambda_1\lambda_2}\to {\cal
L}_{\lambda_1\lambda_2\lambda_3\lambda_4}$, etc. The
corresponding generalization of the equations written down below
is quite obvious.

The equations for the tadpoles are obtained by substitution of
Eqs.~(\ref{delta_g}) and (\ref{delta_d}) (or (\ref{delta_d0}))
into Eqs.~(\ref{Kph}) and (\ref{Kpp}). The final equation for the
K$^{ph}$ tadpole,  in the obvious short notation,  has the form:
\bea K^{ph}_{12} = \sum_{3,4} \int \frac{d\eps}{2\pi i}
\frac{d\omega}{2\pi i}\; \delta {\cal {F}}_{1234}(\omega)
{\cal L}_{34}(\eps ,\omega) g_{34}D_{L}(\omega) + \notag \\
 \sum_{3,4}  {\cal {F}}_{1234} \int \frac{d\eps} {2\pi i}
 \frac{d\omega}{2\pi i} (\delta {\cal{L}} g )_{34} (\eps , \omega )D_{L}(\omega)   + \notag \\
 \sum_{3,4}  {\cal{F}}_{1234} \int \frac{d\eps} {2\pi i} {\cal{L}}_{34}(\eps , \omega_L )
 K^{ph}_{34}\,.
\label{K_ph} \eea

For the K$^{pp}$ tadpole, let us first use Eq.~(\ref{delta_d0})
for $\delta d$. We find: \bea K^{pp}_{12} = 2\sum_{3,4} \int
\frac{d\eps}{2\pi i}
 \frac{d\omega}{2\pi i} \delta {\cal {F}^{\xi}}_{1234}(\omega)
{\cal O}_{34}(\eps,\omega) g_{34}D_{L}(\omega) + \notag \\
 \sum_{3,4}  {\cal F}^{\xi}_{1234} \int \frac{d\eps} {2\pi i}
 \frac{d\omega}{2\pi i} (\delta {\cal{O}} g )_{34} (\eps , \omega )D_{L}(\omega)  + \notag \\
 \sum_3 \int \frac{d\eps}{2\pi i}
 \frac{d\omega}{2\pi i} \delta^{(2)} {\cal
 {F}^{\xi}}_{1233}(\omega) F_3(\eps)
D_{L}(\omega) + \notag \\
  \sum_{3,4}
  {\cal{F}^{\xi}}_{1234}\int \frac{d\eps} {2\pi i} {\cal{O}}_{34}(\eps , \omega_L ) K^{ph}_{34}\,.
\label{K_pp} \eea

The factor 2 in the first term in Eq.~(\ref{K_pp}) appears due to
the fact that, in the small $d$ approximation, the terms $\delta
{\cal F}^{\xi} {\cal O} g $ and $\delta {\cal F}^{\xi} \delta
F^{1}$ in Eq. (\ref{delta_d0}) are equal to each other.

   \begin{figure}[]
 {\includegraphics [height=20mm]{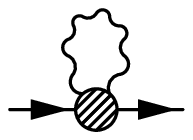}} =
 {\includegraphics [height=20mm]{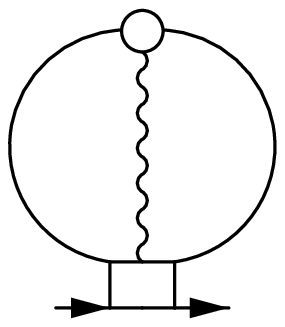}}
 + {\includegraphics [height=20mm]{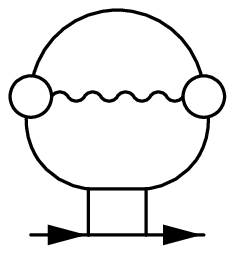}}
+ {\includegraphics [height=20mm]{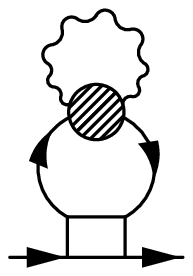}}
 \caption{Equation for the tadpole $K^{ph}$
in non-magic nuclei in the small $d$ approximation. }
\end{figure}

  \begin{figure}[]
  {\includegraphics [height=20mm]{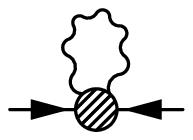}} =
  2 {\includegraphics [height=20mm]{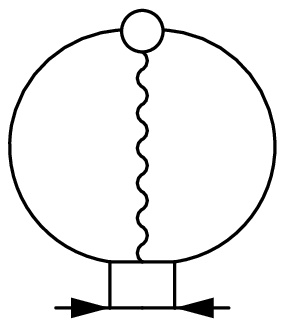}}
 + {\includegraphics [height=20mm]{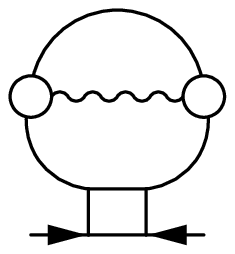}}
 +{\includegraphics [height=20mm]{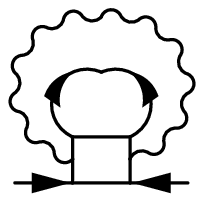}}
+ {\includegraphics [height=20mm]{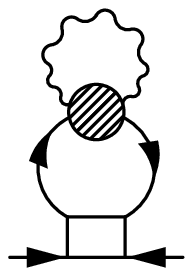}}
 \caption{ Expression for the tadpole $K^{pp}$
in non-magic nuclei in the small $d$ approximation.}
\end{figure}

 In the graphic form, Eqs.~(\ref{K_ph}) and (\ref{K_pp}) are illustrated in
 Fig. 6 and Fig. 7 respectively.
For the sake of simplicity, contrary to Fig. 3, here we do not draw
 all the internal Green functions and omit arrows for
those which are drawn. The arrows are only conserved
 in the cases where it is necessary for understanding. In particular, in the
 last diagrams of both figures the arrows show that we deal
 with the tadpole $K^{ph}$.
  Remember that the second term on the r.h.s. of
Eq.~(\ref{K_ph}) in the detailed presentation includes 8
particular diagrams, in accordance with Eq.~(\ref{d_Lg}). In the
case of $K^{pp}$ tadpole, the number of diagrams is even larger.
Therefore the detailed diagram representation of Eqs.~(\ref{K_ph})
and (\ref{K_pp}) is rather  complicated.

Eq.~(\ref{K_pp}) is convenient for the graphical representation,
but it possesses one drawback: it contains the term with
$\delta^{(2)} {\cal {F}^{\xi}}$ with the ``hidden'' tadpole
$K^{ph}$. To separate the latter one explicitly it is necessary to use
Eq.~(\ref{delta_d}) instead of (\ref{delta_d0}). In the result, we
find: \bea K^{pp}_{12} =
 \sum_{3,4}
\int \frac{d\eps}{2\pi i}
 \frac{d\omega}{2\pi i} (\delta_{L} {\cal F}_{21})_{1234}(\omega)
{\cal L}_{34}(\eps ,\omega) g_{34}D_{L}(\omega) + \notag \\
\sum_{3,4} \int \frac{d\eps}{2\pi i}
 \frac{d\omega}{2\pi i} \delta {\cal {F}^{\xi}}_{1234}(\omega)
{\cal O}_{34}(\eps ,\omega) g_{34}D_{L}(\omega) + \notag \\
\sum_{3,4}  ({\cal F}_{12})_{1234} \int \frac{d\eps} {2\pi i}
 \frac{d\omega}{2\pi i} (\delta {\cal{L}} g )_{34} (\eps , \omega )D_{L}(\omega)  + \notag \\
 \sum_{3,4}  {\cal F}^{\xi}_{1234} \int \frac{d\eps} {2\pi i}
 \frac{d\omega}{2\pi i} (\delta {\cal{O}} g )_{34} (\eps , \omega )D_{L}(\omega)  + \notag \\
  \sum_{3,4}\left[ ({\cal F}_{21})_{1234}\int \frac{d\eps} {2\pi i} {\cal{L}}_{34}(\eps , \omega_L )+
  {\cal{F}^{\xi}}_{1234}\int \frac{d\eps} {2\pi i} {\cal{O}}_{34}(\eps , \omega_L )\right] K^{ph}_{34}\,.
\label{K_pp1} \eea

One remark should be made concerning the integrals over $\omega$
in the above equations for the tadpoles. The poles of the
$D$-function should be taken into account only because they lead
to the terms which strongly depend on the low-laying phonon
frequency $\omega_L$ and other phonon characteristics. They could
change significantly from one nucleus to  another. On the other
hand, the terms appearing due to poles of ${\cal L}$, ${\cal O}$
and other two-particle propagators do not practically depend on
$\omega_L$. They are smooth functions of all the variables under
consideration and should be included into the corresponding
mean-field quantities.

The solution of the integral equations (\ref{K_ph}) and
(\ref{K_pp}) (or (\ref{K_pp1})) yields the tadpole values which,
according to Eqs.~(2) and (3), should be added to the usual
non-local terms. Note that in the approach discussed there is no
need for any new parameters. Below we consider some applications
of the results obtained.

\subsection{Applications to description of the single-particle
characteristics of non-magic nuclei}
In Refs.~\cite{avekaev1999,avekaevlet1999} the approach to
describe the the single-particle strength distribution for
non-magic odd nuclei and  to take into account the phonon
contributions to the single-particle energies and gap values
 has been developed on the basis of generalization of
the Eliashberg theory \cite{eliashberg1961} to nuclei,  with the first
application of the Eliashberg theory to nuclei made  in
\cite{kadmluk1989}.
 The general set of equations of Refs.~\cite{avekaev1999,avekaevlet1999}
 for the energy and gap operators, with account for the dynamic spread of a
 single-particle level (the ``dynamic''
case),   has been derived in the diagonal approximation for the
mass and gap operators.  (Arguments in favor of such an approximation
could be found in \cite{avekaev1999,avekaevlet1999}).
 The equations are
 as follows :
 \bea \eps_{\lambda \eta} = \frac
{\eps_{\lambda}^{(0)} + M^{\rm even}_{\lambda}(E_{\lambda \eta})}
                     {1 + q_{\lambda \eta}(E_{\lambda \eta})}\,, \notag \\
\Delta_{\lambda \eta} = \frac { \Delta_{\lambda}^{(0)} + M^{(1)}_{\lambda} (E_{\lambda \eta})}
{1 + q_{\lambda \eta}(E_{\lambda \eta})}\,,  \notag \\
E_{\lambda \eta} = \sqrt {\eps_{\lambda \eta}^{2} +
\Delta_{\lambda \eta}^{2}}\,, \label{din1} \eea
 where
\beq q_{\lambda \eta} = - \frac {M^{\rm odd}_{\lambda}(E_{\lambda
\eta})}{E_{\lambda \eta}}\,. \label{q} \eeq

Here $M^{\rm even}$ and $M^{\rm odd}$ are even and odd in energy
components of the non-local mass operator $M$ ($M = M^{\rm even} +
M^{\rm odd}$), which enters the r.h.s. of Eq.~(\ref{mig}), and
$M^{(1)}$ is  the same as in Eq.~(\ref{mig1}). The subscript
$\eta$ numerates solutions of the set of
Eqs.(\ref{din1}),(\ref{q}). This yields the distribution of the
single-particle strength in non-magic nuclei.

In order to obtain the single-particle energies and gap values
(the ``static'' case) from Eqs.~(\ref{din1}),(\ref{q}), it is
necessary, for each $\lambda$, to separate the dominant solution
$\eta$ from the set $\{\lambda\eta\}$. For this aim, the
spectroscopic factors should be analyzed. They are given
 \cite{avekaev1999} with
\beq
 S^{\pm}_{\lambda \eta} = \frac {(1 + q_{\lambda \eta})(E_{\lambda \eta} \pm \eps_{\lambda \eta})}
 {\dot {\Theta} _{\lambda}(E_{\lambda \eta})}\,,
 \label{s-factor}
 \eeq
where \beq \Theta_{\lambda}(\eps) = (\eps - \eps^{(0)}_{\lambda} -
M_{\lambda}(\eps))
 (\eps + \eps^{(0)}_{\lambda} + M_{\lambda}^{h}(\eps)) - (\Delta^{(0)}_{\lambda} + M^{(1)}_{\lambda}(\eps))^{2}.
\label{teta}
 \eeq
The $\eta$-component with the maximal spectroscopic factor  should
be associated with the experimental single-particle level, the
details see in \cite{avekaev1999}. Let us denote the observed
single-particle energies and gap values  as $\eps_{\lambda}$ and
$\Delta_{\lambda}$ and the corresponding mean field values as $
\eps^{(0)}_{\lambda}$ and $\Delta^{(0)}_{\lambda}$. They are
related to each other by Eqs.~(\ref{din1}),(\ref{q}) with $\eta$
equal to the dominant value.  Let us rewrite them explicitly,
omitting the subscript $\eta$: \bea \eps_{\lambda} = \frac {
\eps^{(0)}_{\lambda} + M^{\rm even}_{\lambda}(E_{\lambda})}
                     {1 + q_{\lambda}(E_{\lambda })}\,, \notag \\
\Delta_{\lambda} = \frac { \Delta_{\lambda}^{(0)} +
M^{(1)}_{\lambda} (E_{\lambda})}{1 + q_{\lambda }(E_{\lambda})}\,, \notag \\
E_{\lambda} = \sqrt {\eps_{\lambda}^{2} + \Delta_{\lambda
}^{2}}\,, \label{stat1} \eea
 where
\beq q_{\lambda} = - \frac {M^{\rm
odd}_{\lambda}(E_{\lambda})}{E_{\lambda}} \label{statq}\,. \eeq
The energies $\eps_{\lambda}$ and $ \eps_{\lambda}^{(0)}$ are
reckoned from the corresponding chemical potentials $\mu$ and $
\mu^{(0)}$. Note that in Refs.~\cite{avekaev1999,avekaevlet1999}
the phenomenological Saxon-Woods potential was utilized as the
mean field one and the phenomenological pairing forces were used
as well. As far as both of them are adjusted to the observed
values of $\eps_{\lambda}$ and $\Delta_{\lambda}$, a special
``refinement'' procedure mentioned above is necessary to find
$\eps^{(0)}_{\lambda}$ and $\Delta^{(0)}_{\lambda}$ values. It is
described in detail in the cited articles.

Now it is necessary to modify these results in order to include
the tadpoles in accordance with  Eqs.~(\ref{mig}), (\ref{mig1}).
In fact, there was  no specialization of the mass operators in
\cite{avekaev1999,avekaevlet1999} to derive  the relations
(\ref{din1}) and (\ref{stat1}). For this reason, in order to
include the tadpoles into consideration we should just change the
   mass and gap operators of Refs.~\cite{avekaev1999,avekaevlet1999} to the ones
 from Eqs. (\ref{mig}) and
(\ref{mig1}). Remember that the tadpole terms K$^{ph}$ and K$^{pp}$
do not depend on the energy. Supposing, just as the
non-local operators $M,M^{(1)}$, that they  are diagonal in $\lambda$, we
obtain \bea \eps_{\lambda \eta} = \frac { \eps^{(0)}_{\lambda} +
M^{\rm even}_{\lambda}(E_{\lambda \eta})}
                     {1 + q_{\lambda \eta}(E_{\lambda \eta})} + \frac {K^{ph}_{\lambda}}
                     {1 + q_{\lambda \eta}(E_{\lambda \eta})}\,,
                      \notag \\
\Delta_{\lambda \eta} =
 \frac {\Delta^{(0)}_{\lambda} + M^{(1)}_{\lambda} (E_{\lambda \eta})}{1 + q_{\lambda \eta}(E_{\lambda \eta})}
 + \frac {K^{pp}_{\lambda}}
 {1 + q_{\lambda \eta}(E_{\lambda \eta})}\,,
 \notag \\
E_{\lambda \eta} = \sqrt {\eps_{\lambda \eta}^{2} +
\Delta_{\lambda \eta}^{2}}\,, \label{din1new} \eea
 with
 \beq
q_{\lambda \eta} = - \frac {M^{\rm odd}_{\lambda \eta}(E_{\lambda
\eta})}{E_{\lambda \eta}}\,, \label{qnew} \eeq instead of
Eqs.~(\ref{din1}),(\ref{q}). In the same way, instead of
Eqs.~(\ref{stat1}),(\ref{statq}) for the single-particle and gap
values, we find: \bea \eps_{\lambda} = \frac {\eps^{(0)}_{\lambda}
+ M^{\rm even}_{\lambda}(E_{\lambda})}
                     {1 + q_{\lambda}(E_{\lambda})} + \frac {K^{ph}_{\lambda}}
                     {1 + q_{\lambda}(E_{\lambda})}\,,
                      \notag \\
\Delta_{\lambda} = \frac {\Delta^{(0)}_{\lambda} + M^{(1)}_{\lambda}
(E_{\lambda})}{1 + q_{\lambda}(E_{\lambda})}
 + \frac {K^{pp}_{\lambda}}
 {1 + q_{\lambda}(E_{\lambda})}\,,
 \notag \\
E_{\lambda} = \sqrt {\eps_{\lambda}^{2} + \Delta_{\lambda}^{2}}\,,
\label{statnew} \eea
 with
 \beq
q_{\lambda} = - \frac {M^{\rm
odd}_{\lambda}(E_{\lambda})}{E_{\lambda}}\,. \label{q1new} \eeq

 We  see that  both the single-particle energy and gap values are changed due to inclusion
 of the tadpoles, both in the dynamic and static cases. In the latter case,  the
 solution of the set of Eqs.~(\ref{statnew}) should answer the question about
 the total phonon contribution, including the tadpole terms,
to the pairing gap, as compared with the mean field, or
''refined``, value  $\Delta^{(0)}_{\lambda}$. Up to now, all
calculations of the phonon corrections to the gap have ignored the
tadpole  contributions.

 Let us briefly discuss the situation in nuclei without pairing. In this case,
the equations of Sect.~II A for the phonon corrections to the
single-particle energies were solved in the coordinate
representation in \cite{KhS} (see references therein, in
particular, \cite{platonov1980}). It turned out that the tadpole
contribution to $\eps_{\lambda}$ was, as a rule, significant and
comparable with that of the non-local term of the mass operator.
For a qualitative analysis, we limit ourselves with the diagonal
in $\lambda$ approximation, using Eqs.~(\ref{din1new}) and
(\ref{statnew}) without any pairing contribution. For the
spread of a single-particle state we have:
 \beq
\eps_{\lambda \eta} = \eps^{(0)}_{\lambda} +
M_{\lambda}(\eps_{\lambda \eta}) + K^{ph}_\lambda\,, \label{magic}
\eeq
 and for the single-particle energies:
 \beq
\eps_{\lambda} =  \eps^{(0)}_{\lambda} +
M_{\lambda}(\eps_{\lambda}) + K^{ph}_\lambda\,. \label{statmagic}
\eeq As far as K$^{ph}$ does not depend on the energy, the shift
of the solutions, for a fixed $\lambda$, will be the same for all the
values of $\eta$. For the same reason, that is independence of
K$^{ph}$ on energy, the spectroscopic factors which are
determined by the residues of the Green function and, therefore,
by the energy derivative of the   mass operator in
Eq.~(\ref{mig}), are not changed.

  This conclusion about the role of the ph-tadpole in magic nuclei
agrees with the results of calculations for single-particle level
properties in odd neighbors of $^{208}$Pb in Ref.
\cite{litvaring} cited above, where the tadpole contribution was
not taken into account.
 Indeed, as it can be seen from Table III of \cite{litvaring}, the authors obtained a good agreement
 with the experiments for the spectroscopic factors, where there is no tadpole contribution,
 but the agreement  for the single-particle energies is considerably worse
 due to the fact that in this case the tadpole contribution does exist.

Things are different  in nuclei with pairing. In this
case, in the absence of the tadpole, the single-particle spectroscopic
factors are given with Eqs.~(\ref{s-factor}),(\ref{teta}). If the
tadpole terms $K^{ph}$ and $K^{pp}$ are included, as it can be
easily checked, in addition to a change of $E_{\lambda \eta}$ and
$\eps_{\lambda \eta}$, the expression (\ref{s-factor}) itself is
modified. Thus, for non-magic nuclei both the energies and
spectroscopic factors should be changed due to inclusion of the
tadpoles.

\section{Conclusion}
In this work, a consistent approach is developed to include the phonon coupling
in the $g^{2}$ approximation for mass and gap operators in
non-magic nuclei  with explicit consideration of the
tadpole terms, in addition to the usual non-local terms. The general
set of equations for the phonon corrections under discussion is
obtained, which doesn't include any new parameters besides
those used in the self-consistent calculation of the ``zero'' mean
field, i.e. the ones without phonon contributions. This set is
simplified for the case of corrections induced by low-lying
surface phonons in the ``small $d$ approximation''
($d^{(1,2)}<<g$). This approximation  means that the admixture of
pp-phonons with  ph-phonons under consideration, i.e. the
contribution of the $d^{(1,2)}$-vertexes in comparison with the
$g$-vertex, could be neglected . The closed
integral equation for the ph-tadpole $K^{ph}$ as well as  the
integral relation for the pp-tadpole $K^{pp}$ in terms of $K^{ph}$
are obtained. Even for such a simplified case the relations obtained are
much more complicated than those for magic nuclei.

As an application of the relations obtained, the role of the
phonon tadpoles in single-particle strength distribution, in the
single-particle energies and gap values is analyzed. Relations of
Refs.~\cite{avekaev1999,avekaevlet1999}, where only usual
non-local mass operators (in the $g^{2}$ approximation) have been
taken into account, are modified with the explicit inclusion of
the tadpole terms. The set of equations obtained is analyzed. Even
before numerical calculations, the analysis of the structure of
these equations and their comparison with those for magic nuclei,
lead us to a conclusion that the tadpole terms should change
significantly the nuclear characteristics under consideration.
Indeed, on the one hand, this comparison shows that the ph-tadpole
$K^{ph}$ in non-magic nuclei should be close to that in magic
ones. On the other hand, the expression for the pp-tadpole
$K^{pp}$ in terms of $K^{ph}$ shows that the first one has no
smallness in comparison with the second one. Therefore, we could
rely  on the experience of calculations in Ref.~\cite{KhS} for
magic nuclei where the contribution of the tadpole term, e.g., to
the single-particle energies is significant. Note that, contrary
to magic nuclei, in non-magic ones the tadpoles should also change
the  spectroscopic factors. A preliminary analysis of the modified
gap equation shows that here the tadpole could be significant,
too. This is important for the problem of pairing nature  in
finite nuclei.

\section{Acknowledgment}
We thank Prof. S. Krewald for valuable discussions, I. Surkova for
her careful reading of the manuscript and M. Doering for his help
in drawing the Feynman diagrams. The work was partly supported by
the DFG and RFBR grants Nos.GZ:432RUS113/806/0-1 and 05-02-04005,
by the Grant NSh-3004.2008.2 of the Russian Ministry for Science
and Education and by the RFBR grants 06-02-17171-a and
07-02-00553-a.

\end{document}